\documentstyle[12pt]{article}
\newcommand{\be}[1]{\begin{equation} \label{(#1)}}
\newcommand{\ee}{\end{equation}}
\newcommand{\ba}[1]{\begin{eqnarray} \label{(#1)}}
\newcommand{\ea}{\end{eqnarray}}
\newcommand{\nn}{\nonumber}
\newcommand{\rf}[1]{(\ref{(#1)})}

\def\pmb#1{\setbox0=\hbox{#1}%
  \kern-.015em\copy0\kern-\wd0
  \kern.03em\copy0\kern-\wd0
  \kern-.015em\raise.0233em\box0 }
\def \znbb {0\nu\beta\beta}

\begin{document}

\begin{center}

{\bf New Low-energy Leptoquark Interactions}

\bigskip

{M. Hirsch, H.V. Klapdor-Kleingrothaus and S.G. Kovalenko$^*$
\bigskip

{\it
Max-Planck-Institut f\"{u}r Kernphysik, P.O. 10 39 80, D-69029,
Heidelberg, Germany}

\bigskip

$^*${\it Joint Institute for Nuclear Research, Dubna, Russia}
}
\end{center}

\begin{abstract}
We discuss an extension of the standard model (SM) with vector and
scalar leptoquarks. The renormalizable leptoquark Lagrangian consistent
with the SM gauge symmetry is presented including the leptoquark-Higgs
interactions previously not considered in the literature.
We discuss the importance of these new interactions for
leptoquark phenomenology. After the electro-weak symmetry
breaking they generate non-trivial leptoquark mass matrices.
These lead to mixing between different $SU(2)_L$-multiplets
of the leptoquarks and induce at low energies new effective
4-fermion lepton-quark vertices. The latter affect the standard
leptoquark phenomenology. We discuss constraints on these interactions
from the helicity-suppressed $\pi\rightarrow \nu + e$ decay.
\end{abstract}
%%%%%%%%%%%%%%%%%%%%%%%%%%%%%%%%%%%%%%%%%%%%%%%%%%%%%%%%%%%%%%%%%%%%
\bigskip
%\section{Introduction}
%\section{Lagrangian}
%\section{Mass Matrices}
%\section{Effective 4-fermion Interactions}
%\section{Constraints from the Pion Decay}

%\section{Introduction}
The interest on leptoquarks (LQ) \cite{ps} has been renewed
during the last few years since ongoing collider experiments have
good prospects for searching these particles \cite{Lagr1}.
LQs are vector or scalar particles carrying both lepton and baryon numbers
and, therefore, have a well distinguished experimental signature.

LQs can be quite naturally introduced in the low-energy theory as
a relic of a more fundamental theory at some high-energy scale.
In such a way LQs can emerge from grand unified theories (GUT) \cite{GUT},
\cite{sstr}, including the superstring-inspired versions of GUT \cite{sstr},
models of extended technicolour \cite{tech}-\cite{pi_decay} and
composite models \cite{compos}.

The low-energy LQ phenomenology has received considerable attention.
Possible LQ manifestations in various processes have been extensively
investigated \cite{Lagr1}-\cite{bw86}. Various constraints on  LQ masses
and couplings have been deduced from existing experimental data and
prospects for the forthcoming experiments have been estimated.

Direct searches of LQs as s-channel resonances in deep inelastic
ep-scattering at HERA experiments \cite{HERA} placed lower limits on
their mass $M_{LQ} \geq 140-235$GeV \cite{H1-ZEUS}
depending on the LQ type and couplings. With larger accumulated luminosity
HERA will be able to cover almost the whole kinematical region in the
LQ masses up to 296 GeV, for couplings to quarks and leptons above $10^{-2}$.
There are also bounds from
other collider experiments. The LEP experiments exclude any LQ lighter
then 45 GeV \cite{LEP}, the D0 collaboration rules out LQs lighter
than 133 GeV if they couple to the first generation fermions \cite{D0},
and the CDF collaboration sets a corresponding
lower bound at 113 GeV \cite{CDF}.

Dramatic improvements of these constraints are expected in future
experiments at pp \cite{pp}, ep \cite{Lagr1}, \cite{ep},
$e^+ e^-$ \cite{e+e-},\cite{Lagr2}
$e\gamma$ \cite{e-gamma} and $\gamma\gamma$ \cite{gamma-gamma} colliders.

However, at present the most stringent limits on LQs come from
low-energy experiments \cite{DBC}, \cite{Leurer}.
Effective 4-fermion interactions, induced by virtual LQ exchange
at energies much smaller than their masses, can contribute to
atomic parity violation, flavour-changing neutral current (FCNC) processes,
meson decays, meson-antimeson mixing and some rare processes. For instance,
a typical bound on non-chirally coupled LQs imposed by the helicity-suppressed
$\pi\rightarrow e\nu$ decay is $ M^2_{LQ}/|g_L g_R|\geq (100 \mbox{TeV})^2$,
where $g_{L,R}$ are LQ couplings \cite{pi_decay}.

%\section{Lagrangian}

To consider LQ phenomenology in a model-independent fashion one
usually follows some general principles in constructing the Lagrangian
of the LQ interactions with the standard model (SM) fields.
Generic principles are renormalizability  {\bf (p1)}
and invariance {\bf (p2)} under
the SM gauge group $SU(3)_c\otimes SU(2)_L\otimes U(1)_Y$.
In order to obey the stringent constraints from
{\bf (c1)} helicity-suppressed $\pi\rightarrow e\nu$ decay \cite{pi_decay},
from {\bf (c2)} FCNC processes \cite{bw86}, \cite{ps} and
{\bf (c3)} proton stability \cite{bw86}, the following
three assumptions are also commonly adopted:
{\bf (a1)} LQ couplings are "chiral", i.e. each type of LQs couples
either to left-handed or to right-handed quarks only
(call them left- and right-type LQ);
{\bf (a2)} LQ couplings are generation "diagonal", i.e. they couple only to
a single generation of leptons and a single generation of quarks;
{\bf (a3)} LQ interactions conserve baryon (B) and lepton (L) numbers.

Of course, if these empirical assumptions (a1)-(a3) really determine
the LQ interactions, they have to be explained in terms of an underlying
theory predicting light LQs.
Such an explanation can be found for instance, within
theories with a "horizontal" symmetry \cite{Leurer95}.
We will show, however, that assumption (a1) does not solve problem (c1)
since the LQ couplings with the SM Higgs doublet reintroduce
the non-chiral interaction terms.
Therefore, to obey (c1) one should not only claim chirality of
the LQ-quark couplings (a1) but also absence of some LQ-Higgs couplings.
It is unlikely that both requested properties can have the same origin
in the underlying theory.

In the following we consider changes in LQ phenomenology caused by
the LQ-Higgs interactions. We base our consideration on the general
principles (p1),(p2) as well as on the assumptions (a1)-(a3).

In the literature only LQ-lepton-quark interaction terms
have been considered.
They have the following form \cite{Lagr1} for the scalar LQs
\ba{S-l-q}
{\cal L}_{S-l-q} &=&
\lambda^{(R)}_{S_0}\cdot \overline{u^c} P_R e \cdot S_0^{R\dagger} +
%%%%%%%%%%%%%%%%%%%%%%%%%%%%%%%%%%%
\lambda^{(R)}_{\tilde S_0}\cdot \overline{d^c} P_R e \cdot
\tilde{S}_0^{\dagger} + \\ \nn
%%%%%%%%%%%%%%%%%%%%%%%%%%%%%%%%%%%
&+& \lambda^{(R)}_{S_{1/2}}\cdot \overline{u} P_L l
\cdot {S}_{1/2}^{R\dagger} +
%%%%%%%%%%%%%%%%%%%%%%%%%%%%%%%%%%%
\lambda^{(R)}_{\tilde S_{1/2}}\cdot \overline{d} P_L l \cdot
\tilde{S}_{1/2}^{\dagger} + \\ \nn
%%%%%%%%%%%%%%%%%%%%%%%%%%%%%%%%%%%
&+& \lambda^{(L)}_{S_0}\cdot \overline{q^c} P_L i \tau_2 l \cdot
S_0^{L\dagger} +
%%%%%%%%%%%%%%%%%%%%%%%%%%%%%%%%%%%
\lambda^{(L)}_{S_{1/2}}\cdot \overline{q} P_R i \tau_2 e \cdot
S_{1/2}^{L\dagger} + \\ \nn
%%%%%%%%%%%%%%%%%%%%%%%%%%%%%%%%%%%
&+& \lambda^{(L)}_{S_1}\cdot \overline{q^c} P_L i \tau_2
\hat{S}_1^{\dagger} l + h.c.
\ea
and for the vector LQs
\ba{V-l-q}
{\cal L}_{V-l-q} &=&
\lambda^{(R)}_{V_0}\cdot \overline{d} \gamma^{\mu} P_R e
\cdot V_{0\mu}^{R\dagger} +
%%%%%%%%%%%%%%%%%%%%%%%%%%%%%%%%%%%
\lambda^{(R)}_{\tilde V_0}\cdot \overline{u} \gamma^{\mu} P_R e \cdot
\tilde{V}_{0\mu}^{\dagger} + \\ \nn
%%%%%%%%%%%%%%%%%%%%%%%%%%%%%%%%%%%
&+& \lambda^{(R)}_{V_{1/2}}\cdot \overline{d^c} \gamma^{\mu} P_L l
\cdot {V}_{1/2\mu}^{R\dagger} +
%%%%%%%%%%%%%%%%%%%%%%%%%%%%%%%%%%%
\lambda^{(R)}_{\tilde V_{1/2}}\cdot \overline{u^c} \gamma^{\mu} P_L l \cdot
\tilde{V}_{1/2\mu}^{\dagger} + \\ \nn
%%%%%%%%%%%%%%%%%%%%%%%%%%%%%%%%%%%
&+&
\lambda^{(L)}_{V_0}\cdot \overline{q} \gamma^{\mu} P_L l \cdot
V_{0\mu}^{L\dagger} +
%%%%%%%%%%%%%%%%%%%%%%%%%%%%%%%%%%%
\lambda^{(L)}_{V_{1/2}}\cdot \overline{q^c} \gamma^{\mu} P_R e \cdot
V_{1/2\mu}^{L\dagger} + \\ \nn
%%%%%%%%%%%%%%%%%%%%%%%%%%%%%%%%%%%
&+& \lambda^{(L)}_{V_1}\cdot \overline{q} \gamma^{\mu}  P_L
\hat{V}_{1\mu}^{\dagger} l + h.c.
\ea
Here $P_{L,R} = (1\mp\gamma_5)/2$;  $q$ and $l$ are the quark
and the lepton doublets; $S_i^{j}$ and $V_i^{j}$
are the scalar and vector LQs with the weak isospin i=0, 1/2, 1
coupled to left-handed ($j = L$) or right-handed ($j = R$)
quarks respectively.
The LQ quantum numbers are listed in Table 1.
For LQ triplets $\Phi_1 = S_1, V_1^{\mu}$ we use the notation
$\hat{\Phi}_1 = \vec\tau\cdot\vec{\Phi}_1$.

%%%%%%%%%%%%%%%%%%%%%%%%%%%%%%%%%%%%%%%%%%%%%%%%%%%%%%%%%%%%%%%%%
\begin{table}[t]
{Table 1:
The Standard model assignments of the
scalar ${ S}$ and vector ${ V}_{\mu}$  leptoquarks (LQ).
($Y = 2(Q_{\it em} - T_3)$)
}\\[5mm]
%\vspace*{5mm}
\begin{tabular}{|c|c|c|c|c|}\hline
LQ & $SU(3)_c$ & $SU(2)_L$ &    $Y$    & $Q_{\it em}$ \\
\hline
%&&&&\\
$ S_0$         & ${\bf 3}$ & ${\bf 1}$ & -2/3 &  -1/3\\
\hline
%&&&&\\
$\tilde{S}_0$   & ${\bf 3}$ & ${\bf 1}$ & -8/3 &  -4/3\\
\hline
%&&&&\\
$S_{1/2}$    & ${\bf 3^\ast}$ & ${\bf 2}$ & -7/3  & (-2/3, -5/3)\\
\hline
%&&&&\\
$\tilde S_{1/2}$   & ${\bf 3^\ast}$ & ${\bf 2}$ & -1/3  & (1/3, -2/3)\\
\hline
%&&&&\\
$S_1$    & ${\bf 3}$ & ${\bf 3}$ & -2/3  & (2/3, -1/3,-4/3)\\
\hline
%%%%%%%%%%%%%%%%%%%%%%%%%%%%%%%%%%%%%%%%%%%%%%%%%%%%%%%%%%%
%&&&&\\
$ V_0$         & ${\bf 3^\ast}$ & ${\bf 1}$ & -4/3 &  -2/3\\
\hline
%&&&&\\
$\tilde{V}_0$   & ${\bf 3^\ast}$ & ${\bf 1}$ & -10/3 &  -5/3\\
\hline
%&&&&\\
$V_{1/2}$    & ${\bf 3}$ & ${\bf 2}$ & -5/3  & (-1/3, -4/3)\\
\hline
%&&&&\\
$\tilde V_{1/2}$   & ${\bf 3}$ & ${\bf 2}$ & 1/3  & (2/3, -1/3)\\
\hline
%&&&&\\
$V_1$    & ${\bf 3^\ast}$ & ${\bf 3}$ & -4/3  & (1/3, -2/3,-5/3)\\
\hline
\end{tabular}
\end{table}
%%%%%%%%%%%%%%%%%%%%%%%%%%%%%%%%%%%%%%%%%%%%%%%%%%%%%%%%%%%%%%%%%%%%%

There are no fundamental reasons forbidding LQ interactions with the
standard model Higgs doublet H.
The most general form of the LQ-Higgs interaction Lagrangian,
consistent with (p1) and (p2) can be expressed as
\ba{LQ-Higgs}
{\cal L}_{LQ-H} &=& h^{(i)}_{S_0} H i\tau_2 \tilde{ S}_{1/2}\cdot
S_0^{i} +
h^{(i)}_{V_0} H i\tau_2 \tilde{ V}_{1/2}^{\mu}\cdot V_{0\mu}^{i}
+ \\ \nn
&+& h_{S_1} H i\tau_2 \hat S_{1}\cdot \tilde{S}_{1/2} +
h_{V_1} H i\tau_2 \hat V_{1}^{\mu}\cdot \tilde{V}_{1/2\mu} +\\ \nn
&+& Y_{S_{1/2}}^{(i)} \left(H i\tau_2 S_{1/2}^{i}\right)\cdot
\left(\tilde{S}_{1/2}^{\dagger} H\right) +
Y_{V_{1/2}}^{(i)} \left(H i\tau_2  V_{1/2}^{\mu(i)} \right)\cdot
\left(\tilde{V}_{1/2\mu}^{\dagger} H\right) + \\ \nn
&+& Y_{S_1} \left(H i\tau_2 { \hat S}_{1}^{\dagger} H\right)\cdot
\tilde{S}_{0} +
Y_{V_1} \left(H i\tau_2  \hat V_{1\mu}^{\dagger} H\right)\cdot
\tilde{V}_{0}^{\mu} + \\ \nn
&+& \kappa_{S}^{(i)} \left(H^{\dagger} \hat S_{1} H\right)\cdot
S_{0}^{i\dagger} +
\kappa_{V}^{(i)} \left(H^{\dagger} \hat V_1^{\mu} H\right)\cdot
V_{0\mu}^{i\dagger} + h.c.
-
\\ \nn
&-& \left(\eta_{\Phi} M^2_{\Phi} -
g_{\Phi}^{(i_1i_2)} H^{\dagger} H\right) \Phi^{i_1\dagger} \Phi^{i_2}.
\ea
Here $H =  \mbox{$ \left( \begin{array}{cc}
        H^+\\
        H^0
        \end{array} \right) $}$ is the SM $SU(2)_L$-doublet Higgs field.

$\Phi^{i}$ is a cumulative notation for all leptoquark fields with
$i = L,R$ (the same for $i_{1,2}$).
We included diagonal
mass terms $\eta_{\Phi} M^2_{\Phi}\Phi^{\dagger}\Phi$ of the scalar
($\eta_S = 1$) and the vector ($\eta_V = -1$) LQ fields.
These terms can be generated by spontaneous breaking of the underlying
symmetry down to the electro-weak gauge group at some high-energy scale.

%\section{Mass Matrices}

The subsequent electro-weak symmetry breaking at the Fermi scale
produces additional non-diagonal LQ mass terms leading to
non-trivial mixing between LQs from different $SU(2)_L$ multiplets
as well as between left- and right-types of LQs, discussed above.

The relevant LQ mass matrices can be read off from  eq. \rf{LQ-Higgs}.

There are 8 non-diagonal LQ mass matrices squared (I=S,V):
\ba{M_k}  %28
%%%%%%%%%%%%%%%%%%%%%%%%%%%%%%%%%%%%%%%%%%%%%%%%%%%%%%%%
{\cal M}_I^2(Q_I^{(1)})    &=& \eta_I\cdot\left(
                        \begin{array}{cccc}
\eta_I \bar{M}_{I_0^L}^2 & g_{I_0}^{LR}|v|^2 & h_{I_0}^{(L)} v
& \kappa_{I}^{(L)} |v|^2  \\
g_{I_0}^{LR}|v|^2  & \eta_I \bar{M}_{I_0^R}^2 & h_{I_0}^{(R)} v
& \kappa_{I}^{(R)} |v|^2 \\
h_{I_0}^{(L)} v  & h_{I_0}^{(R)} v
&\eta_I \bar{M}_{\tilde{I}_{1/2}}^2 & h_{I_1} v  \\
\kappa_{I}^{(L)} |v|^2 &\kappa_{I}^{(R)} |v|^2 &
h_{I_1} v &\eta_I \bar{M}_{I_1}^2\\
\end{array}
                     \right),\\   \label{M_2}
%%%%%%%%%%%%%%%%%%%%%%%%%%%%%%%%%%%%%%%%%%%%%%%%%%%%%%%%
{\cal M}_I^2(Q_I^{(2)})  &=& \eta_I\cdot\left(
                        \begin{array}{cccc}
 \eta_I\bar{M}_{\tilde{I}_{1/2}}^2 & Y_{I_{1/2}}^L v^2
&Y_{I_{1/2}}^R v^2 & \sqrt{2} h_{I_1} v \\
Y_{I_{1/2}}^L v^2  & \eta_I\bar{M}_{I_{1/2}^L}^2
& g_{I_{1/2}}^{(LR)} |v|^2  & 0\\
  Y_{I_{1/2}}^R v^2  & g_{I_{1/2}}^{(LR)} |v|^2
& \eta_I\bar{M}_{I_{1/2}^R}^2 & 0  \\
\sqrt{2} h_{I_1} v  & 0 & 0 & \eta_I\bar{M}_{I_1}^2  \\
 \end{array}
                     \right),  \\    \label{(M_3)}
%%%%%%%%%%%%%%%%%%%%%%%%%%%%%%%%%%%%%%%%%%%%%%%%%%%%%%%%
{\cal M}_I^2(Q_I^{(3)})
&=& \eta_I\cdot\left(
                        \begin{array}{cc}
 \eta_I\bar{M}_{\tilde{I}_0}^2 & \sqrt{2} Y_{I_1}v^2 \\
 \sqrt{2} Y_{I_1}v^2 & \eta_I\bar{M}_{I_1}^2 \\
 \end{array}
                     \right), \\                  \label{(M_4)}
%%%%%%%%%%%%%%%%%%%%%%%%%%%%%%%%%%%%%%%%%%%%%%%%%%%%%%%%
{\cal M}_I^2(Q_I^{(4)})
&=& \eta_I\cdot\left(
                        \begin{array}{cc}
 \eta_I\bar{M}_{I_{1/2}^L}^2 & -g_{I_{1/2}}^{(LR)} |v|^2 \\
 -g_{I_{1/2}}^{(LR)} |v|^2 & \eta_I\bar{M}_{I_{1/2}^R}^2 \\
 \end{array}
                     \right),
\ea
where $\bar{M}_I^2 = M_I^2 + \eta_I g_I |v|^2$  is the "shifted" diagonal
mass, $v^2 = <H^0>^2 = (2\sqrt{2} G_F)^{-1}$ is the SM Higgs field
vacuum expectation value, $G_F$ is the Fermi constant and $\eta_{S,V} = 1,-1$.
The cumulative notation $ {\cal M}_I^2(Q_I^{(k)})$ encodes the mass matrices
squared for the scalar (I=S) and vector ($I=V_{\mu}$) LQs with electric charges
$Q_S^{(1)} = Q_V^{(2)} = -1/3, \ \ Q_S^{(2)} = Q_V^{(1)} = -2/3, \ \
Q_S^{(3)} = Q_V^{(4)} = -4/3, \ \  Q_V^{(3)} = Q_S^{(4)} = -5/3$
in the interaction eigenstate basises:
(k=1) $I(Q_I^{(1)}) = (I_0^L, I_0^R, \tilde{I}_{1/2}^{\dagger},I_1)$;
(k=2) $I(Q_I^{(2)}) = (\tilde{I}_{1/2}, I_{1/2}^L, I_{1/2}^R, I_1^{\dagger})$;
(k=3) $I(Q_I^{(3)}) = (\tilde{I}_0, I_1)$;
(k=4) $I(Q_I^{(4)}) = (I_{1/2}^L, I_{1/2}^R)$.
Thus, there is a non-trivial mixing of LQs from different $SU(2)_L$ multiplets
as well as the  $I^L-I^R$ mixing. The latter spoils chirality of
the LQ-quark-lepton couplings (a1) and leads to reappearance of the problem
with the constraint (c1).

The mass matrices of all other LQ fields remain diagonal
after electro-weak symmetry breaking.

To obtain
observable predictions from the LQ-lepton-quark interaction Lagrangian
in eqs. \rf{S-l-q}-\rf{V-l-q}, fields with
non-diagonal mass matrices
have to be rotated to the mass eigenstate basis $I'$. This can be done
in the standard way
\ba{eigen}
I(Q) = {\cal N}^{(I)}(Q)\cdot I'(Q)
\ea
where ${\cal N}^{(I)}(Q)$
are
orthogonal matrices such that
${\cal N}^{(I)T}(Q_I)\cdot {\cal M}_I^2(Q)\cdot {\cal N}^{(I)}(Q)
= Diag\{M_{I_n}^2\}$ with the $M_{I_n}$ being the mass of the relevant
mass eigenstate field $I'$.
All phenomenological consequences of the LQ interactions in
eqs. \rf{S-l-q}-\rf{LQ-Higgs} should be derived in terms of these fields $I'$.

%\section{Effective 4-fermion Interactions}

In this letter we concentrate on the LQ induced 4-fermion lepton-quark
effective interactions. For their derivation one has to substitute the
expression \rf{eigen} to eqs. \rf{S-l-q}-\rf{V-l-q} and
"integrate out"
heavy LQ fields $I'$.
For vanishing LQ-Higgs couplings (eq. \rf{LQ-Higgs}) these
interaction terms are listed in ref. \cite{DBC}.
Mixing between different $SU(2)_L$ multiplets of LQs leads to new terms,
vanishing in the limiting case of decoupled LQ and Higgs sectors.
Below we list only those new terms which can be most stringently restricted
from low-energy processes\footnote{A detailed study of phenomenological
implications of the LQ-Higgs couplings in eq. \rf{LQ-Higgs}
will be given elsewhere.}. After Fierz rearrangement they take the form
\ba{mix_terms} \nn
{\cal L}_{mix}^{eff} &=&
(\bar\nu P_R e^c)
\left[\frac{\epsilon_S}{M_S^2} (\bar u P_R d) +
\frac{\epsilon_V}{M_V^2} (\bar u P_L d)\right]
%%%%%%%%%%%%%%%%%%%%%%%%%%%%%%%%%%%%%%%%%%%%%%%%%%%%
+ (\overline{\nu^c} P_L e^c)
\left[\frac{\omega_S}{M_S^2} (\bar u P_L d) +
\frac{\omega_V}{M_V^2} (\bar u P_R d)\right] \\
&-& (\bar\nu \gamma^{\mu} P_L e^c)
\left[
\left(\frac{\alpha_S^{(R)}}{M_S^2}  + \frac{\alpha_V^{(R)}}{M_V^2} \right)
(\bar u \gamma_{\mu} P_R d) -
\sqrt{2}\left(\frac{\alpha_S^{(L)}}{M_S^2}  + \frac{\alpha_V^{(L)}}{M_V^2} \right)
(\bar u \gamma_{\mu} P_L d)\right],
%-\\ \nn
%&-& (\overline{\nu^c} \gamma^{\mu} P_R e^c)
%\left[\omega_L (\bar u \gamma_{\mu} P_L d) -
%\omega_R (\bar u \gamma_{\mu} P_R d)\right]
\ea
where
\ba{coeff} \nn
\epsilon_I &=&
2^{-\eta_I}
\left[\lambda_{I_1}^{(L)}\lambda_{\tilde{I}_{1/2}}^{(R)}
\left(\theta_{43}^I(Q_I^{(1)}) + \eta_I \sqrt{2} \theta_{41}^I(Q_I^{(2)})\right) -
\lambda_{I_0}^{(L)}\lambda_{\tilde{I}_{1/2}}^{(R)}
\theta_{13}^I(Q_I^{(1)})\right], \\ \nn
%%%%%%%%%%%%%%%%%%%%%%%%%%%%%%%%%%%%%%%%%%%%%%%%%%%%
\omega_I &=&
2^{-\eta_I}
\left[\lambda_{I_0}^{(L)}\lambda_{I_0}^{(R)}\theta_{12}^I(Q_I^{(1)})
+ \lambda_{I_0}^{(R)}\lambda_{I_1}^{(L)} \theta_{42}^I(Q_I^{(1)})
+ \lambda_{I_{1/2}}^{(L)}\lambda_{I_{1/2}}^{(R)} \theta_{32}^I(Q_I^{(2)})
\right], \\
%%%%%%%%%%%%%%%%%%%%%%%%%%%%%%%%%%%%%%%%%%%%%%%%%%%%%
\alpha_I^{(L)} &=&
\frac{2}{3 + \eta_I} \lambda_{I_{1/2}}^{(L)}\lambda_{I_1}^{(L)}
\theta_{24}^I(Q_I^{(2)}), \ \ \
\alpha_I^{(R)} = \frac{2}{3 + \eta_I}
\lambda_{I_0}^{(R)} \lambda_{\tilde{I}_{1/2}}^{(R)}
\theta_{23}^I(Q_I^{(1)}).
\ea
Here we introduced a mixing parameter
\ba{mp}
\theta_{kn}^I(Q) = \sum_{l} {\cal N}_{kl}^{(I)}(Q) {\cal N}_{nl}^{(I)}(Q)
\left(\frac{M_I}{M_{I_l}(Q)}\right)^2,
\ea
where $Q = -1/3,-2/3$ and $I=S,V$. Common mass scales $M_S$ of scalar
and $M_V$ of vector LQs were introduced for convenience.

%\section{Constraints from the Pion Decay}

The interaction terms eq. \rf{mix_terms} contribute to various
low-energy processes.
Using existing experimental data one can obtain constraints on the
relevant coupling constants. Here we are not going to discuss
this subject in detail but rather present only the most stringent bounds
from the helicity-suppressed decay $\pi\rightarrow e\nu$.
This process is especially sensitive to the first two scalar-pseudoscalar
terms leading to a helicity-unsuppressed amplitude.
Assuming no spurious cancellations between different contributions we
derive on the basis of ref. \cite{DBC} the following severe constraints:

\ba{pion}
\epsilon_I, \omega_I\leq
5\times 10^{-7}\left(\frac{M_I}{100\mbox{GeV}}\right)^2
\ea
Other couplings in eq. \rf{mix_terms} are much weaker constrained by
low-energy processes previously considered in connection with
the LQ phenomenology \cite{DBC}, \cite{Leurer}.
We expect that new stringent constraints on LQ couplings can be derived from
neutrinoless double beta decay ($\znbb$).
The first and the last terms with $\epsilon$, $\alpha$ couplings
in eq. \rf{mix_terms} contribute to
this exotic process within the conventional mechanism based on
Majorana neutrino exchange between decaying nucleons. Because of the
specific helicity structure the corresponding amplitude acquires
an enormously
large enhancement factor $p_F/m_{\nu}\sim 10^8$ ($p_F\approx 100$MeV
is the Fermi momentum and $m_{\nu}$ is the neutrino mass)
compared to the standard charged current contribution.
As a result non-observation of $\znbb$-decay casts stringent constraints
on the LQ parameters.
This subject will be considered in a separate paper.

In conclusion, we have pointed out that the interactions of leptoquarks with
the standard model Higgs field modify low-energy leptoquark phenomenology.
They generate new 4-fermion lepton-quark couplings which contribute
to various low-energy processes. We have considered as an example the
helicity-suppressed $\pi\rightarrow e\nu$ decay,
which stringently constraints special combinations of
the leptoquark couplings including couplings to the Higgs field
(see eq. \rf{pion}).
We stress that an underlying high-energy scale theory containing
light leptoquarks
must explain not only the chirality of leptoquark couplings
to quarks and leptons  (see (a1) at the beginning)
but also the absence (or smallness) of at least those leptoquark-Higgs
couplings which are suppressed by the constraints in eq. \rf{pion}.

\bigskip
\centerline{\bf ACKNOWLEDGMENTS}
We thank V.A.~Bednyakov for helpful discussions.
M.H. would like to thank the Deutsche Forschungsgemeinschaft for financial
support by grants kl 253/8-1 and 446 JAP-113/101/0.

\end{document}